# Energy states of the Hulthen plus Coulomb-like potential with position-dependent mass function in external magnetic fields

M. Eshghi*,[1], R. Sever[2], S.M. Ikhdair[3,4]

[1] *Young Researchers and Elite club, Central Tehran Branch, Islamic Azad University, Tehran, Iran*

[2] *Department of Physics, Middle East Technical University, Ankara, Turkey*

[3] *Department of Physics, Faculty of Science, An-Najah National University, Nablus, West Bank, Palestine*

[4] *Department of Electrical Engineering, Near East University, Nicosia, Northern Cyprus, Mersin 10, Turkey*

## Abstract

We need to solve a suitable exponential form of the position-dependent mass (PDM) Schrödinger equation with a charged particle placed in the Hulthen plus Coulomb-like potential field and under the influence of the external magnetic and Aharonov-Bohm (AB) flux fields. The bound state energies and their corresponding wave functions are calculated for spatially-dependent mass distribution function of a physical interest. A few plots of some numerical results to the energy are shown.

**Keywords:** Schrödinger equation; Hulthen plus Coulomb-like potential; position-dependent mass distribution functions; perpendicular magnetic and Aharonov-Bohm flux fields.

## 1. Introduction

The exact solutions of quantum wave equations, expressed in analytical form, describing one-electron atoms and few-body systems are essential in studying the atomic structure theory and more areas. In fact, the exact analytical solutions are essentially used in quantum-chemical, quantum electrodynamics and theory of molecular vibrations. They are also used to examine the correctness of models, approximations in computational physics, nuclear physics, nanostructures and computational chemistry [1-16].

On the other hand, the position-dependent mass (PDM) idea arises after the effect of the periodic field on the non-relativistic motion of electrons in periodic lattices. In fact, it happens in typical semiconductors by the effect of impurities in perturbed periodic lattices [17]. Recently, a considerable interest, in the mass dependence on the inter-nuclear distance, has been revived in solving different equations with various physical potential models [18-29]. Furthermore, a number of works takes the effect of an electric or magnetic fields into account in studying different systems [30-34]. For further study we cite Refs. [35,36]. In addition, it is found more interesting that nearly all desired analytic solutions of the non-relativistic equation have been expressed in terms of hypergeometric functions [38-40].

---

* ***Email corresponding author:*** *eshgi54@gmail.com; m.eshghi@semnan.ac.ir*

However, in all these areas, the study of the non-relativistic and relativistic quantum dynamics of charged particles in the presence of magnetic fields and Aharonov-Bohm (AB) flux fields which are perpendicular to the plane where the particles are confined, has been carried out over the past years [41]. In fact, the investigation of systems consisting of non-relativistic as well as relativistic charged particles, that are confined to the magnetic fields has attracted much attention due to their applications (such as in graphene [15, 16, 42], semiconductor structures [43], chemical physics [44], molecular vibrational and rotational spectroscopy of molecular physics [45], biology [46], environmental sciences [47], and cosmic string [48]), Recently, Eshghi et al [49] have solved the Schrödinger equation with the superposition of Morse-plus-Coulomb potentials with two different physically PDM distribution functions of the exponential and inverse-square forms under the external perpendicular magnetic and AB flux fields. Also, these authors have investigated the Schrodinger equation with a position-dependent mass charged particle interacted via the superposition of the Morse-plus-Coulomb potentials under the influence of external magnetic and Aharonov–Bohm flux fields [49]. However, in the present work, we intend to extend the work in Ref. [50] and solve the Schrödinger equation with a charged particle under the Hulthen plus Coulomb-like potential field having the general form:

$$V(\rho) = -\frac{V_0 e^{-\lambda\rho}}{1 - qe^{-\lambda\rho}} + \frac{V_1}{\rho}, \qquad (1)$$

where $\lambda$ is a mass constant, $V_0$ and $V_1$ are positive potential parameters and $q$ is the real parameter with physically presumed PDM distribution function of the exponential form:

$$M(\rho) = \frac{M_0}{1 - qe^{-\lambda\rho}}, \qquad (2)$$

where $\lambda$ is a mass constant being placed in the external perpendicular magnetic and AB flux fields.

In Figures 1 to 4, we show plots for the potential form $V(\rho)$ in (1) and mass distribution function $M(\rho)$ in (2) versus $\rho$ for various parameters values.

For example, in Fig. 1, we show the behavior of the potential model $V(\rho)$ as changing versus $\rho$ for $q = 3$ and $q = 4$, namely, $q > 0$. Further, Fig. 2 shows the behavior at $V(\rho)$ as changing with $\rho$ for $q = -1, -2$ and $-3$, namely $q < 0$, for $M_0 = 0.4, 0.5, 0.6$.

For $q > 0$, Fig. 3 shows the mass function versus $\rho$ for $q = 3$ and $q = 4$, and for case of $q < 0$, Fig. 4 shows the mass function versus $\rho$ for $M_0 = 0.4, 0.5, 0.6$.

The organization of this paper is as follows. In Section 2, we solve the Schrodinger equation for a charged particle under the influence of Hulthen plus Coulomb-like potential with a suitable choice of spatially dependent mass function and subjects to the external magnetic fields. The series method is used to determine the energy states and their corresponding wave functions. Further, we use the energy levels to obtain the thermodynamic quantities of the system in a systematic manner. Section 3 is devoted for our discussions and conclusions.

## 2. Schrödinger Equation with q-Deformed PDM Function

In this section, we seek to solve the Schrödinger equation for a charged particle with a physically position-dependent mass (PDM) distribution function, interacted via the Hulthen plus Coulomb-like potential field and exposed to external perpendicular magnetic and AB flux fields treated in two-dimensional space cylindrical coordinates. We seek to calculate the bound state energies and their corresponding wave functions. The general form of the Schrodinger equation for a charged particle with PDM system under the influence of a certain potential field and in the presence of the vector potential is given by

$$\left(\hat{p} + \frac{e}{c}\vec{A}\right) \cdot \frac{1}{2M(\rho)} \left(\hat{p} + \frac{e}{c}\vec{A}\right) \Psi(\rho, \varphi, z) = [E_{nm} - V(\rho)]\Psi(\rho, \varphi, z), \qquad (3)$$

where $M(\rho)$ and $\vec{A}$ is the spatially dependent mass and mass constant, respectively.

We assume that the vector potential has the simple form: $\vec{A} = \left(0, \frac{B_0 e^{-\lambda\rho}}{1-qe^{-\lambda\rho}} + \frac{\Phi_{AB}}{2\pi\rho}, 0\right)$ where $B_0$ and $\Phi_{AB}$ are the magnetic and AB flux fields.

After having substituted the vector potential (1) and the spatially dependent mass (2) into (3) and using the approximation of Aldrich [51] with some lengthy but straightforward calculations, we arrive at the following radial second-order differential equation:

$$\frac{d^2R(\rho)}{d\rho^2} + \frac{(1+q)\lambda e^{-\lambda\rho}}{1-qe^{-\lambda\rho}}\frac{dR(\rho)}{d\rho}$$

$$+ \left[\vartheta(m,\lambda,\Phi_{AB},B_0)\frac{e^{-2\lambda\rho}}{(1-qe^{-\lambda\rho})^2} + \delta(\lambda,V_0,V_1,M_0)\frac{e^{-\lambda\rho}}{(1-qe^{-\lambda\rho})^2}\right. \tag{4}$$

$$\left. - \omega(m,\Phi_{AB})\frac{e^{-\lambda\rho}}{1-qe^{-\lambda\rho}} + \varepsilon(M_0,E_{nm})\frac{1}{1-qe^{-\lambda\rho}}\right]R(\rho) = 0,$$

and

$$\vartheta(m,\lambda,\Phi_{AB},B_0) = -m^2\lambda^2 - \frac{2meB_0\lambda}{\hbar c} - \left(\frac{eB_0}{\hbar c}\right)^2 - \left(\frac{e}{\hbar c}\right)^2\frac{B_0\lambda\Phi_{AB}}{\pi} - \left(\frac{e\lambda\Phi_{AB}}{2\pi\hbar c}\right)^2,$$

$$\delta(\lambda,V_0,V_1,M_0) = \frac{2M_0}{\hbar^2}(V_0 - V_1\lambda), \quad \omega(m,\Phi_{AB}) = \frac{em\Phi_{AB}}{\pi\hbar c}, \quad \varepsilon(M_0,E_{nm}) = \frac{2M_0}{\hbar^2}E_{nm}.$$

where $m$ is the magnetic quantum number.

Now, using the parameterized Nikoforov-Uvarov method [52] with the following substitution $z = \frac{1}{1-qe^{-\lambda\rho}}$ into Eq. (4), we can simply write Eq. (4) as

$$\frac{d^2F(z)}{dz^2} + \frac{(1-q)z}{qz(1-z)}\frac{dF(z)}{dz} + \frac{1}{z^2(1-z)^2}[-\xi_2 z^2 + \xi_1 z - \xi_0]F(z) = 0, \tag{5}$$

with the following list of identifications:

$$-\xi_2 = \frac{1}{q\lambda^2}\left(\vartheta(m,\lambda,\Phi_{AB},B_0) - \delta(\lambda,V_0,V_1,M_0)\right),$$

$$\xi_1 = \frac{1}{q\lambda^2}\left(-2\vartheta(m,\lambda,\Phi_{AB},B_0) - \delta(\lambda,V_0,V_1,M_0) - \omega(m,\Phi_{AB}) + \varepsilon(M_0,E_{nm})\right)$$

$$-\xi_0 = \frac{1}{q\lambda^2}\left(\vartheta(m,\lambda,\Phi_{AB},B_0) + \omega(m,\Phi_{AB})\right),$$

$$c_1 = 1, \quad c_2 = 2 - \frac{1}{q}, \quad c_3 = 1, \quad c_4 = 0, \quad c_5 = -\frac{1}{2q}, \tag{6}$$

$$c_6 = \frac{1}{4q^2} + \xi_2, \quad c_7 = -\xi_1, \quad c_8 = \xi_0,$$

$$c_9 = \frac{1}{4q^2} + \xi_2 - \xi_1 + \xi_0, \quad c_{10} = 1 - 2\sqrt{\xi_0},$$

$$c_{11} = 2 + 2\sqrt{\frac{1}{4q^2} + \xi_2 - \xi_1 + \xi_0} - 2\sqrt{\xi_0},$$

$$c_{12} = -\sqrt{\xi_0}, \quad c_{13} = -\frac{1}{2q} - \sqrt{\frac{1}{4q^2} + \xi_2 - \xi_1 + \xi_0} + \sqrt{\xi_0},$$

we can write the energy eigenvalues in a rather simpler form as

$$E_{nm} = \frac{1}{2W_3}[-(W_1 + 2n)^2 + 2(n^2 + W_1 n + W_4)]$$

$$\pm \sqrt{\frac{(W_1 + 2n)^2 - 2(n^2 + W_1 n + W_4)}{4W_3} - \frac{(n^2 + W_1 n + W_4)^2 - W_2(W_1 + 2n)^2}{W_3^2}}, \quad (7)$$

where we have defined

$$W_1 = 1 - 2\sqrt{\frac{1}{q\lambda^2}\left(-\vartheta(m, \lambda, \Phi_{AB}, B_0) - \omega(m, \Phi_{AB})\right)},$$

$$W_2 = \frac{2}{q\lambda^2}\delta(\lambda, V_0, V_1, M_0) + \frac{1}{4q^2},$$

$$W_3 = \frac{2M_0}{q\lambda^2 \hbar^2}, \quad (8)$$

$$W_4 = -\sqrt{\frac{1}{q\lambda^2}\left(-\vartheta(m, \lambda, \Phi_{AB}, B_0) - \omega(m, \Phi_{AB})\right)} + \frac{1}{2q}$$

$$+ \frac{1}{q\lambda^2}[\delta(\lambda, V_0, V_1, M_0) - \omega(m, \Phi_{AB})].$$

Here we need to examine the behavior of the energy states in Eq. (7), so we plot the energy state as a function of magnetic field and AB flux field for various mass parameter considering the positive sign in Eq. (7) as shown in Figures 5 and 6.

Therefore, it is obvious from Fig. 5 that the energy state is sensitive to the changing magnetic field and it becomes strongly bound with stronger field and as we change the mass parameter $M_0$. The energy state is overlapping for different values of mass parameter. Therefore, there is a limit on taking the mass of a charged particle for different field strengths. Clearly, the taken mass value must be small enough to avoid any overlapping of the energy state. In Fig.6, the energy state decreases with increasing the AB flux field, $\Phi$. However, the energy states is higher for larger mass value.

In Figures 7 and 8, we also plot energy state as function of magnetic and AB flux fields (curve and linear, respectively) for the negative sign case in Eq. (7). Clearly, there is no limit on mass parameter value in this case. The energy state curve becomes strongly negative or bound with increasing magnetic field for a chosen mass parameter value. However, the energy state curve is linear and becomes strongly bound with increasing AB flux field.

To proceed in our discussion, we can use energy spectrum formula (7) to study the thermodynamics properties of a charged particle in the presence and absence of scalar and vector fields. Having calculated the energy states, we can immediately obtain the thermodynamics quantities of the system in a systematic manner. In order to obtain all thermodynamic quantities of the nonrelativistic particle system, we should concentrate, at first, on the calculation of the partition function $Z$. In this case, the partition function $Z$ at temperature $T$, is being obtained throughout the Boltzmann factor as $Z = \sum_{n=0}^{\infty} exp(-\beta E_{n,m})$ where $\beta = 1/k_B T$ and $k_B$ is the Boltzmann constant [53].

On the other hand, after using the Eqs. (6) and parameterized NU method [52], we can simply obtain the wave function as

$$F(z) = Nz^{-\sqrt{\frac{1}{q\lambda^2}(-\vartheta(m,\lambda,\Phi_{AB},B_0)-\omega(m,\Phi_{AB}))}}$$
$$\times (1-z)^{\frac{1}{2q}+\sqrt{\frac{1}{4q^2}+\frac{1}{q\lambda^2}(3\delta(\lambda,V_0,V_1,M_0)-\varepsilon(M_0,E_{nm}))}} \quad (9)$$
$$\times P_n^{\left(-2\sqrt{\frac{1}{q\lambda^2}(-\vartheta(m,\lambda,\Phi_{AB},B_0)-\omega(m,\Phi_{AB}))},\ 2\sqrt{\frac{1}{4q^2}+\frac{1}{q\lambda^2}(3\delta(\lambda,V_0,V_1,M_0)-\varepsilon(M_0,E_{nm}))}\right)}(1-2z),$$

where $N$ is the normalized constant.

## 3. Concluding Remarks

We have solved the Schrödinger equation for a charged particle with an exponential form position-dependent mass (PDM) placed in field of the general superposition of Húlthen plus Coulomb-like potential fields under the influence of external magnetic and AB flux fields. We have calculated the bound state energies and the corresponding wave functions with a suitable change to the dependent mass variable as function of the magnetic and AB flux fields by using the parameterized NU method. Finally, some results of the energy values are shown in Figs. 5 to 8.

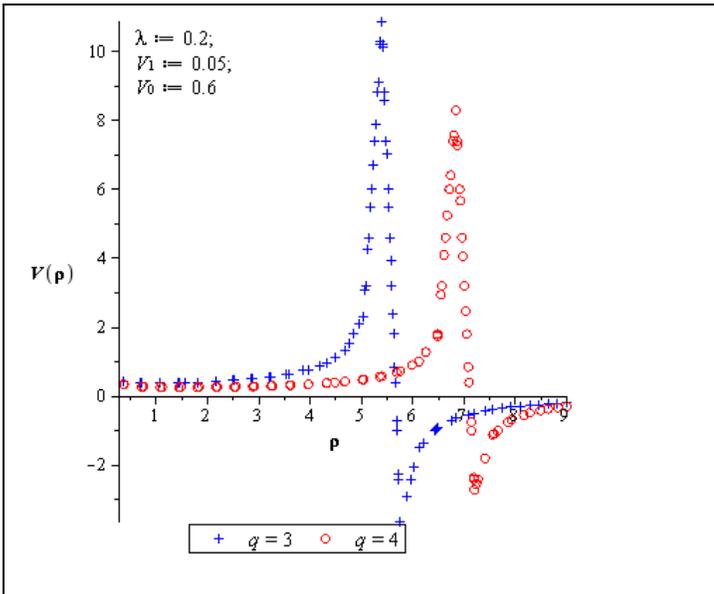

Fig. 1. The potential model (1) versus $\rho$ for two selected values of $q > 0$

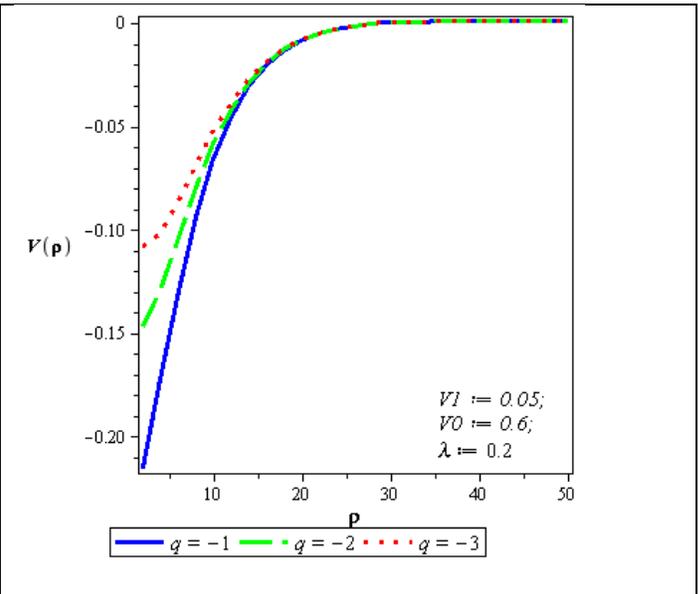

Fig. 2. The potential model (1) versus $\rho$ for three selected values of $q < 0$

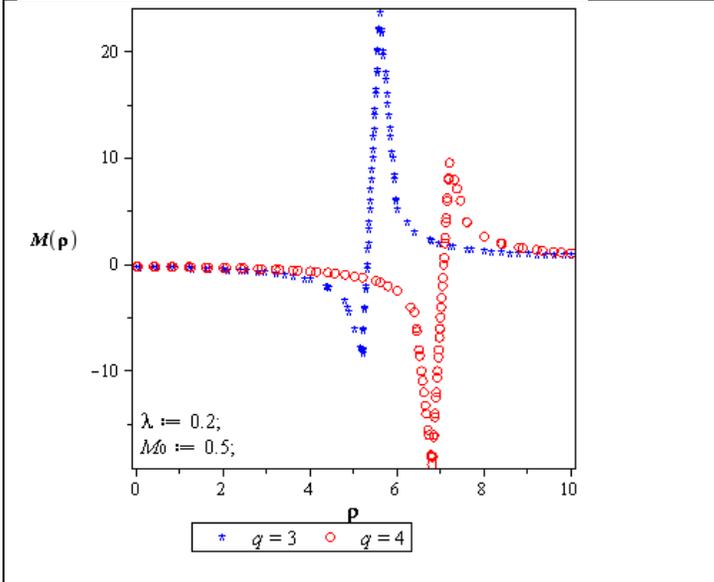

Fig. 3. The mass function (2) versus $\rho$ for $q > 0$ with $M_0 = 0.4, 0.5, 0.6$.

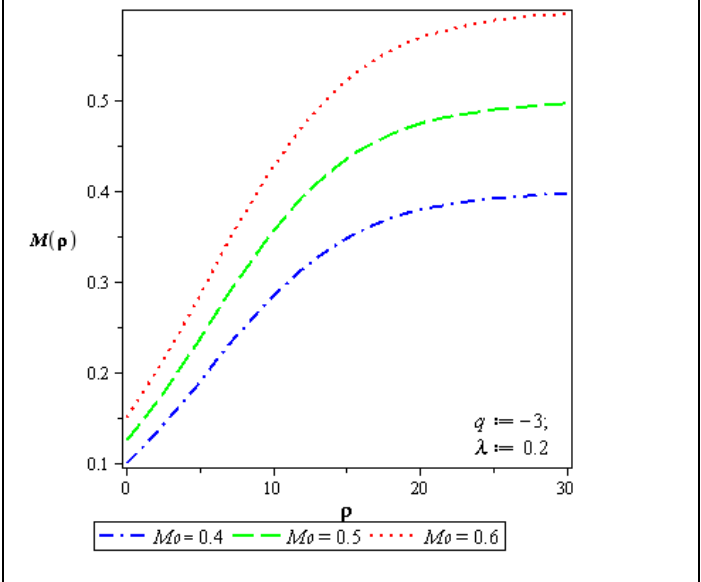

Fig. 4. The mass function (2) versus $\rho$ for $q < 0$ with $M_0 = 0.4, 0.5, 0.6$.

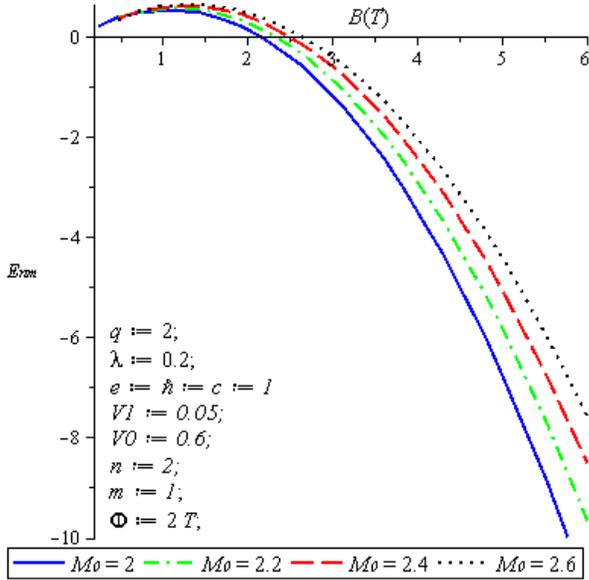

**Fig. 5.** The energy state versus magnetic field, $B$, for various values of $M_0$

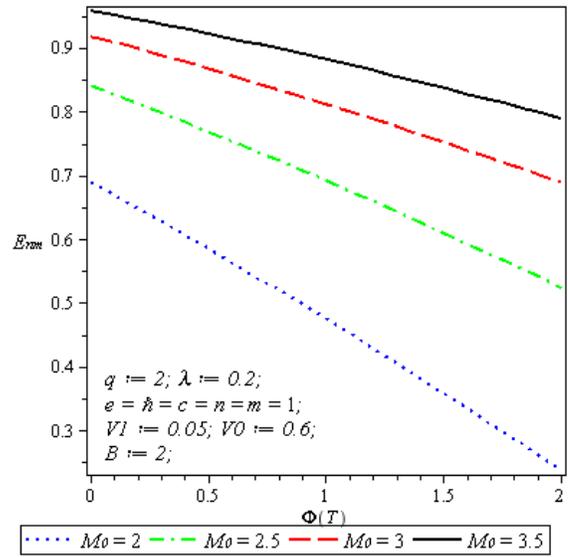

**Fig. 6** The energy state versus the AB flux field, $\Phi$, for various values of $M_0$

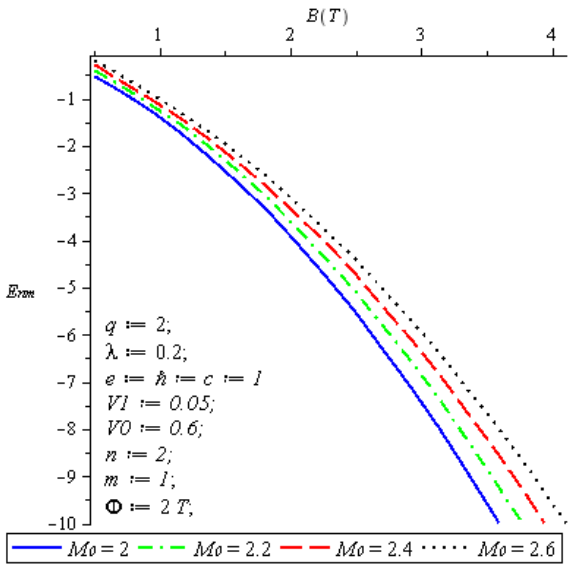

**Fig. 7.** The energy states versus of the magnetic field, $B$, for different values of $M_0$

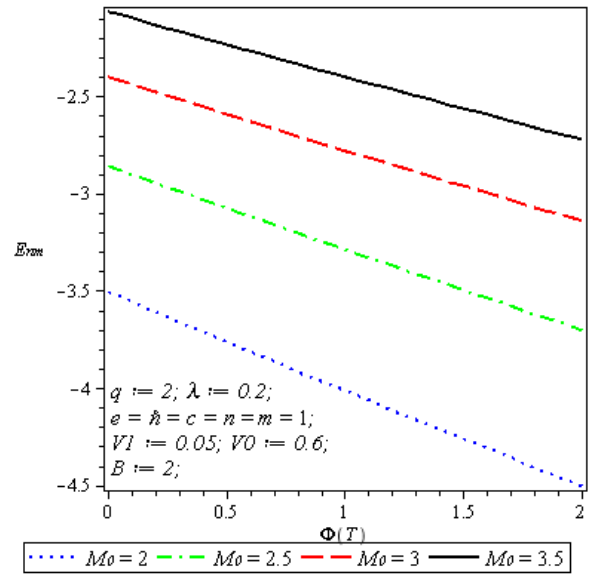

**Fig. 8** The energy states versus of the AB flux field, $\Phi$, for different values of $M_0$